\documentclass[aps, pre, reprint]{revtex4-1}
\usepackage[english]{babel}
\makeatletter
\def\bbl@set@language#1{%
  \edef\languagename{%
    \ifnum\escapechar=\expandafter`\string#1\@empty
    \else\string#1\@empty\fi}%
  \@ifundefined{babel@language@alias@\languagename}{}{%
    \edef\languagename{\@nameuse{babel@language@alias@\languagename}}%
  }%
  \select@language{\languagename}%
  \expandafter\ifx\csname date\languagename\endcsname\relax\else
    \if@filesw
      \protected@write\@auxout{}{\string\select@language{\languagename}}%
      \bbl@for\bbl@tempa\BabelContentsFiles{%
        \addtocontents{\bbl@tempa}{\xstring\select@language{\languagename}}}%
      \bbl@usehooks{write}{}%
    \fi
  \fi}
\newcommand{\DeclareLanguageAlias}[2]{%
  \global\@namedef{babel@language@alias@#1}{#2}%
}
\makeatother
\DeclareLanguageAlias{en}{english}
\usepackage{amsmath,amssymb}
\usepackage{graphicx}
\usepackage{tikz}
\usetikzlibrary{positioning,calc}
\usepackage{color}
\usepackage{siunitx}
\newcommand{\meff}{m_\mathrm{eff}}
\newcommand{\mER}{\hat{m}_\mathrm{ER}}
\newcommand{\mEQ}{\hat{m}_\mathrm{EQ}}
\newcommand{\mLR}{\hat{m}_\mathrm{LR}}
\newcommand{\mNLR}{\hat{m}_\mathrm{NLR}}

\begin{document}
\title{Description of spreading dynamics by microscopic network models and macroscopic branching processes can differ due to coalescence}
\author{Johannes Zierenberg$^{1,2}$}
\author{Jens Wilting$^{1}$}
\author{Viola Priesemann$^{1,2}$}
\author{Anna Levina$^{3,4}$}
\affiliation{
  $^1$ Max Planck Institute for Dynamics and Self-Organization, Am Fassberg 17, 37077 G{\"o}ttingen, Germany,\\
  $^2$ Bernstein Center for Computational Neuroscience, Am Fassberg 17, 37077 G{\"o}ttingen, Germany,\\
  $^3$ University of T\"ubingen, Max Planck Ring 8, 72076 T\"ubingen, Germany,\\
  $^4$ \mbox{Max Planck Institute for Biological Cybernetics, Max Planck Ring 8, 72076 T\"ubingen, Germany}
}
%


\begin{abstract}
  Spreading processes are conventionally monitored on a macroscopic level by
  counting the number of incidences over time. The spreading process can then be
  modeled either on the microscopic level, assuming an underlying interaction
  network, or directly on the macroscopic level, assuming that microscopic
  contributions are negligible. The macroscopic characteristics of both
  descriptions are commonly assumed to be identical. In this work, we show that
  these characteristics of microscopic and macroscopic descriptions can be
  different due to coalescence, i.e., a node being activated at the same time by
  multiple sources. In particular, we consider a (microscopic) branching network
  (probabilistic cellular automaton) with annealed connectivity disorder, record
  the macroscopic activity, and then approximate this activity by a (macroscopic) branching
  process. In this framework, we analytically calculate the effect of
  coalescence on the collective dynamics. We show that coalescence leads to a
  universal non-linear scaling function for the conditional expectation value of
  successive network activity. This allows us to quantify the difference between
  the microscopic model parameter and established macroscopic estimates. To
  overcome this difference, we propose a non-linear estimator that correctly
  infers the model branching parameter for all system sizes.  
\end{abstract}
\maketitle


\section{Introduction}

A multitude of spreading processes are influencing our life. Examples
include the spread of news, opinions, or rumors~\cite{daley1964, moreno2004},
the outbreak of diseases~\cite{pastor-satorras2015, dearruda2018}, the escalation of
economic crises~\cite{garas2010}, or the propagation of spiking activity in neural
systems~\cite{harris1963, munoz2018}. Mathematically, the unifying feature of
these processes is that some signal (infection, information, spike) spreads
through a system. 

Spreading can be modeled on a microscopic node-to-node level, which requires to
make assumptions about the interaction graph and the rules how a signal may
propagate from one node to the next. This typically involves stochastic
processes, such as (probabilistic) cellular automata~\cite{wolfram1983,
domany1984, hinrichsen2000}, contact processes~\cite{harris1974,hinrichsen2000},
or interacting Hawkes processes~\cite{hawkes1971}. In particular, infectious
diseases have been modeled by so-called susceptible-infectious models or
generalizations thereof~\cite{hethcote2000}, whereas spike-propagation in neural
networks has been modeled by so-called branching networks~\cite{haldeman2005,
kinouchi2006, levina2008, larremore2011, larremore2012,
friedman2012, priesemann2014, wilting2018, zierenberg2018}, Hawkes
processes~\cite{chevallier2017,chevallier2018,kossio2018}, or probabilistic
integrate-and-fire networks~\cite{larremore2014, li2019}. These models can be
either constructed as independent-interaction models (static interactions), or
as threshold models with interactions depending on the states of the interacting
partners~\cite{pei2013}. We focus here on independent-interaction models. The
advantage of these microscopic models is that one can directly study the effect
of network topology --- such as country maps~\cite{hufnagel2004, brockmann2013,
matamalas2018}, daily transportation patterns~\cite{belik2011}, social
links~\cite{fowler2008, maier2017}, or connectomes~\cite{moretti2013} --- and
even time-varying networks --- such as diffusive motion of
nodes~\cite{bonabeau2002}.  

Alternatively, spreading can be modeled on a macroscopic population level, assuming either
that there is no explicit network or that network heterogeneity is averaged
out. Macroscopic models describe the development of population or network
activity without assuming a specific network topology, e.g., modeling the
total number of infected people at each time step by simply assuming a general
statistics of how each infected person spreads the disease. Classical examples
include the branching process~\cite{harris1963}, Kesten
process~\cite{kesten1973}, or other processes of the autoregressive
$\text{AR}(n)$ family. These processes are frequently used to describe
spreading dynamics in real-world systems, because parameter estimation for them
has been well established. For example, branching processes have been used to
explain data in neuroscience~\cite{beggs2003, pasquale2008, petermann2009,
das2018a, wilting2018}, epidemics~\cite{ball1995,wilting2018}, or
economics~\cite{vanderlans2009}.

Microscopic and macroscopic spreading models can, however, be quite different:
spreading processes in microscopic models can interact with each other (e.g.,
when a node being already activated by another node), while in macroscopic
models this is typically not the case (e.g., in the branching process each
element generates new descendants independently of the number of descendants of
the other elements). As one of the consequences, the network activity of
microscopic spreading models is upper bounded by the network size, wheres the
population activity of macroscopic spreading models can in principle diverge. On
the other hand, one can find equivalent behavior in the limit of low external
and internal activation (e.g. branching network and branching process, see
below). As a result, macroscopic approximations are often used to describe
typical behavior of microscopic models~\cite{ball1995, haldeman2005,
wilting2018, zierenberg2018}, in part because such approximations offer the
advantage of analytical tractability. However, if one wants to model a real
system, one has to carefully weight the assumptions one has to make in either
microscopic or macroscopic models.
\begin{figure}[t]
  \includegraphics{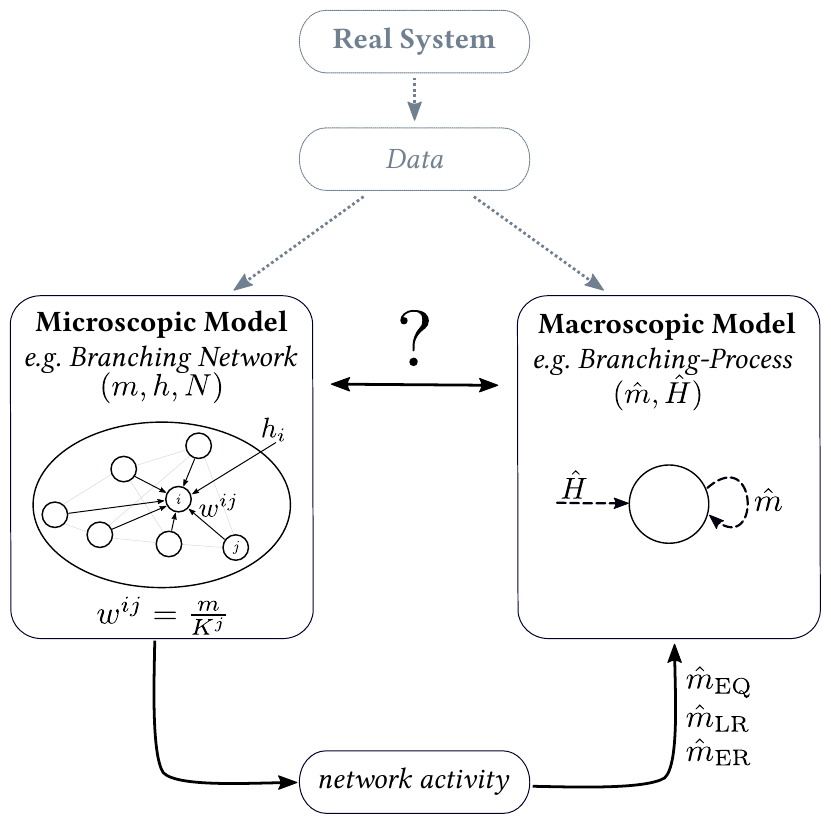}
  \caption{%
    Illustration of the question we address (top) and how we approach it
    (bottom). Question: Given that a real system, for which we measure data
    (e.g. the number of incidences per time step), can be approximated by a
    microscopic model (assuming an underlying interaction network with update
    rules that produce incidences) or a  macroscopic model (on the level of the
    number of incidences), what is the relation between the resulting
    macroscopic dynamics of both models? Approach: We use a branching network
    with annealed connectivity disorder to generate population activity and
    approximate this activity by a branching process. Thereby, we avoid
    systematic errors from (i) an unknown real system and (ii) required
    approximations for the network model. We can thus focus on different
    branching-process approximations ($\mEQ$, $\mLR$, and $\mER$) and can
    compare the resulting macroscopic dynamics of both models.
    \label{figProblem}
  }
\end{figure}

In this work, we address the question ``Given that a real system (and its
measured macroscopic data) can be approximated by a microscopic or a
macroscopic model, what is the relation between the resulting macroscopic
dynamics of both models?'' To approach this question (Sec.~\ref{secModel}), we
simplify the problem by generating the data (network activity) directly with
the microscopic model (branching network), thereby avoiding assumptions about
the real system and how to model this with a microscopic model
(Fig.~\ref{figProblem}). We can thus focus on the approximations in the
macroscopic model (branching process). We reveal analytically and numerically
that conventional estimators of the branching parameter can be biased, i.e.,
estimates do not agree with the microscopic model parameter
(Sec.~\ref{secResultAnaCoalescence}--\ref{secResultFssDriven}). The reason for
this bias is coalescence (the simultaneous activation of one node by multiple
sources). We propose a non-linear estimator that correctly infers the
microscopic model parameter from the network activity
(Sec.~\ref{secResultFssNonLin}). Finally, we discuss our results with
implications for general spreading processes (Sec.~\ref{secDiscussion}).

\section{Model and methods}
\label{secModel}
We use branching networks as a microscopic model and generate macroscopic
observables (network activity). We then approximate these observables by a
branching process as a macroscopic model and compare the spreading rates between
microscopic model and macroscopic approximation (Fig.~\ref{figProblem}). In
this framework, the spreading rate is called branching parameter $m$. To
distinguish parameters on the microscopic and macroscopic level, we denote
macroscopic parameters by a hat, e.g., $\hat{m}$. 

\subsection{Branching network}
Consider a network with $N$ nodes. Time progresses in discrete time steps
$\Delta t$ (here $\Delta t=1$). Each node $i$ can be either silent ($s^i_t=0$)
or active ($s^i_t=1$), thus the (macroscopic) network activity is given by
$A_t=\sum_{i=1}^N s^i_t$. Activation of a node can be induced in two ways. 
First, internally a node $i$ can be activated with probability $w^{ij}$
(connection weight) by another node $j$ that was active in the previous time
step. The connection weight determines the microscopic dynamics. 
Second, a node can be activated by an external Poisson input if one or more
inputs arrive within $\Delta t$. For a Poisson process of rate $h$, the
probability that no input arrives is $\exp(-h\Delta t)$ such that the
activation probability through external input is $\lambda(h)=1-\exp(-h\Delta
t)$. The network-wide external input rate is then $H=hN$. 
An activated node transitions back from $s^j_t=1$ to $s^j_{t+1}=0$ in the
next time step unless activated again, which corresponds to a refractory period
smaller than $\Delta t$ in the modeling of neuronal activity.

The microscopic dynamics are controlled by the branching parameter $m$ ---
motivated by a branching process (see below) --- which requires to construct a
network where a single activation of any node causes on average $m$ active
nodes in the next time step. The simplest way to achieve this is to set
connection weights $w^{ij}=\frac{m}{K^j}$, if $i$ is one of the $K^j$ outgoing
connections from node $j$, and $w^{ij}=0$ otherwise. We here consider the
mean-field scenario of annealed connectivity disorder: connections between
nodes are redrawn in each step with probability $K/N$ and nodes are activated
internally with probability $w=m/K$.  For $K$ sufficiently large, this is
mathematically equivalent to an all-to-all connected network with
$w^{ij}=w=\frac{m}{N}$, including potential self-coupling. This mean-field
connectivity ignores spatial heterogeneity~\cite{lopes2019}, but our results
can be adapted to mean-field approximations for quenched disorder over static
random (Erd\H{o}s-R{\'e}nyi-type) networks~\cite{kinouchi2006}.

A naive implementation of our model dynamics on an all-to-all connected network
is computationally very expensive. This is because for each active node we
would have to draw $N$ random numbers to check for internal activation. To
reduce numerical complexity, we note that in the mean-field case the number of
nodes $k$ activated by a single active node is distributed binomially
$P(k)=\binom{N}{k}w^k(1-w)^{N-k}$. Instead of going over all $N$ connected
nodes, we can thus first draw a number $k$ from the binomial distribution, and
then draw $k$ random nodes without repetitions to be activated. This procedure
is significantly more efficient, especially for large system sizes. For our
finite-size scaling analysis we simulated networks up to size $N=2^{20}\approx 10^6$ for
$10^7$ times steps.

The branching network is in fact a special probabilistic cellular
automaton~\cite{domany1984,kinouchi2006} in the universality class of directed
bond percolation~\cite{hinrichsen2000, henkel2008}.

\subsection{Branching process}
The branching process~\cite{harris1963} is a time-discrete stochastic Markov
process: If at time $t$ there are $A_t$ elements, then at time $t+1$ each of
these elements generates a random integer number of descendants $x_{t}^i$. If
this internal generation of descendants is complemented by random external
input $y_t$, one speaks about a branching process with
immigration~\cite{heathcote1965,pakes1971} --- or a driven branching process.
The number of elements at time $t+1$ can be written as 
\begin{equation}
  A_{t+1} = \sum_{i=1}^{A_t} x_t^i + y_t.
\end{equation}
Many results about branching processes only depend on the average number of
internally generated elements $\hat{m}=\langle x_t^{i}\rangle$, called
branching parameter, and on the average number of externally generated elements
$\hat{y}$ per time step. In order to compare with the branching network, we
identify $\hat{y}=\hat{H}\Delta t$, where $\hat{H}$ is the total rate of the
Poisson-distributed external input. Recall that we use $\hat{m}$ to distinguish
parameters on the macroscopic level from those at a microscopic level. Using
the Markovian nature of branching processes we describe the time evolution of
population activity $A_t$  by the conditional expectation value
\begin{equation}\label{eqBPcondexp}
  \langle A_{t+1} | A_t\rangle =  \hat{m} A_t + \hat{H}\Delta t.
\end{equation}
The branching process thus corresponds to the class of processes with an
autoregressive representation.

\begin{figure*}[]
  \includegraphics{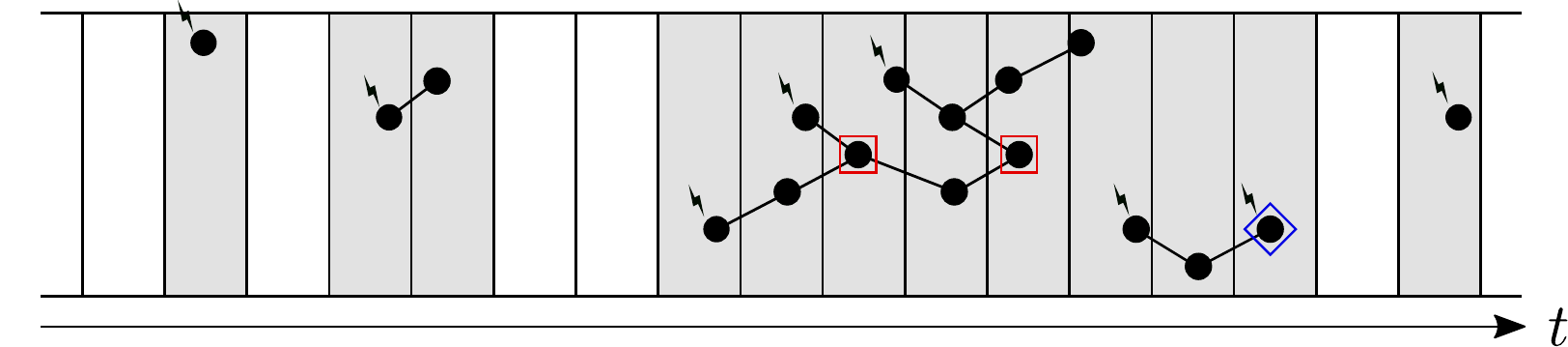}
  \caption{%
    Illustration activity propagation (black dots) and coalescence in branching networks: Different
    realizations of the branching process on the network converge onto each
    other, with the consequence that a node is activated by two or more sources.
    We distinguish internal coalescence (red squares), where two or more source from within the
    network activate the same target , and external coalescence  (red diamond),
    where external input (denoted by black lightnings) contributes. The y-axis
    denotes different nodes, the x-axes time in discrete steps. Time-steps
    without any activity are white, the others are gray.
    \label{figCoalescence}
    }
\end{figure*}

The population activity $A$ of a driven branching process can be calculated in
a mean-field approximation. Assuming a stationary population activity ($\hat{m}<1$),
we can neglect fluctuations and only consider the expectation value over
Eq.~\eqref{eqBPcondexp}. The law of total expectation then implies $A = \langle
\langle A_{t+1}| A_t\rangle \rangle = \hat{m}A+\hat{H}\Delta t$, where the
expectation value of the network activity is the network rate $A=\langle
A_t\rangle$. Solving for the network rate leads to
\begin{equation}\label{eqBPactivity}
  A = \frac{\hat{H}\Delta t}{1-\hat{m}}.  
\end{equation}
which is only well defined for $\hat{m}<1$ and diverges for $\hat{m}\to 1$.

\subsection{Branching process approximation}
The framework of a \textit{branching process} can be used to infer the
branching parameter as a proxy for the spreading dynamics from a time series of
\textit{network} activity $\{A_t\}$. This is clearly an approximation, because
(i) the network sets an upper bound on $A_t$ and (ii) because nodes interact
with each other. Within this approximation there are yet different possible
approaches. We present three in the following. 

The first and easiest approach we consider is to estimate the branching
parameter via Eq.~\eqref{eqBPactivity}. Assuming, the network-wide external
input rate is known from the microscopic dynamics ($\hat{H}=hN$), one can
estimate the branching parameter from the expectation value of the network rate
$\langle A\rangle$ (expected rate = ER) as  
\begin{equation}
  \mER = 1- \frac{\hat{H}\Delta t}{\langle A\rangle}.
\end{equation}
This clearly neglects fluctuations and will therefore be biased for $m\approx
1$ where fluctuations are large. In addition, it is obviously upper-bounded,
$\mER\leq 1$.

The second approach we consider is more elaborate, it is based on the
relationship for the conditional expectation value, Eq.~\eqref{eqBPcondexp}.
Assuming a separation of timescales (STS), i.e., that no external activation is
delivered while the network is active, one can neglect $\hat{H}$ and define the
expectation value of the quotient of subsequent network activity (expected
quotient = EQ)
\begin{equation}\label{eqModelEstimateBP}
  \mEQ = \left\langle\frac{A_{t+1}}{A_t}\right\rangle_{A_t>0}.
\end{equation}
This estimator has been widely applied to neural data~\cite{haldeman2005,
priesemann2014,yu2014}. However, as we will show in more details later, it is
strongly biased for networks subject to non-negligible input rate ($H>0$) as noted
before~\cite{priesemann2014}. 

The third approach we consider is again based on the conditional expectation
value, Eq.~\eqref{eqBPcondexp}, but explicitly considers the presence of external input
rate. In fact, it is well known that the first moments of a driven branching
process can be estimated with a linear regression~\cite{harris1963, wei1990,
wilting2018}:
\begin{eqnarray}\label{eqModelEstimateLR}
  \mLR &=& \frac{\mathrm{Cov}[A_{t+1},A_t]}{\mathrm{Var}[A_t]},\\
  \hat{H}_\mathrm{LR} &=& \frac{1}{\Delta t}\left(\langle A_{t+1}\rangle - \mLR\langle A_t\rangle\right).
\end{eqnarray}
Besides the additional benefit of simultaneously estimating the input rate, this estimator
is invariant to sub-sampling, i.e., when activity is recorded from only (small)
parts of the network~\cite{wilting2018}. For stationary activity, $\langle
A_t\rangle=\langle A_{t+1}\rangle=A$, Eq.~\eqref{eqModelEstimateLR} simplifies to
\begin{equation}
  \mLR = \frac{\langle A_{t+1} A_t\rangle - A^2}{\langle A_t^2\rangle - A^2}.
\end{equation}

We here defined the estimators in terms of expectation values $\langle
O_t\rangle$, formally defined for infinitely long time series of some observable
$O_t$. For a finite time series the expectation values themselves have to be
estimated by the time average
\begin{equation}
  \overline{O_t}=\sum_{t=1}^T
  O_t\underset{T\rightarrow\infty}{\longrightarrow}\langle O_t\rangle. 
\end{equation}

\section{Results}
\label{secResult}

\subsection{Analytic results on effects of coalescence in driven branching networks}
\label{secResultAnaCoalescence}
During spreading dynamics on a finite network, activity of different nodes may
interfere with each other.  With increasing network activity, the probability
increases that a node receives activation from two or more nodes in the same
time step. We call such multiple activations of a node \textit{coalescence}
because different branches of the spreading process coalesce
(Fig.~\ref{figCoalescence}). We further distinguish between \textit{internal
coalescence}, where a node gets activated by two or more nodes from within the
network, and \textit{external coalescence}, where a node gets additionally
activated by external input. 


As a consequence of coalescence, the effective number of internally activated
nodes, or in other words the (microscopic) \textit{effective branching
parameter} $\meff(A_t)$, will be diminished. One may expect that in the limit
$N\to\infty$ the effective branching parameter approaches the model branching
parameter, but we will show that this is not always the case. For driven
branching networks, one can imagine that the external input initiates
independent spreading processes. Already the initiation can cause external
coalescence with the present processes. In addition, the individual spreading
processes interact in the sense of a neutral theory~\cite{martinello2017},
which leads to internal coalescence.

 
In order to derive the effective branching parameter, we first derive the
probability that a given node $i$ is activated. Due to potential coalescence
this is not straightforward, but we can compute the probability
$p^\mathrm{int}_\mathrm{na}$ that node $i$ is not activated by node $j$. For
generality, we consider the annealed disorder with connection selection
probability $K/N$ and activation probability $m/K$. Then,
$p^\mathrm{int}_\mathrm{na}= (1-K/N) + (K/N)(1-m/K)=1-m/N$, which is independent of $K$.
Also, node $i$ is not activated by the external input with probability
$p^\mathrm{ext}_\mathrm{na} = 1-\lambda(h)$. Considering that there are $A_t$
active nodes from which node $i$ can be activated, the probability of activating
node $i$ is 
\begin{align}\label{eqProbActivationDriven}
  P\left[s^i_t=1| A_t,N,m,h\right] 
  &= 1- \left(
  p^\mathrm{int}_\mathrm{na}\right)^{A_t}p^\mathrm{ext}_\mathrm{na}\nonumber\\
  &= 1-\left(1-\frac{m}{N}\right)^{A_t}\left(1-\lambda(h)\right).
\end{align}
Since this holds for any node in the network, we can generalize $P\left[s^i_t=1|
A_t,N,m,h\right]=p(A_t)$.
Then, the probability for network activity $A_{t+1}$, given network activity
$A_t$ in the previous time step, is expressed by the binomial distribution

\begin{equation}\label{eqActivity}
   P\left[A_{t+1} | A_t,N,m,h\right] = \binom{N}{A_{t+1}}
  p(A_t)^{A_{t+1}}\left(1-p(A_t)\right)^{N-A_{t+1}},
\end{equation}
with expectation value 
\begin{equation}\label{eqResultAnaCond}
  \langle A_{t+1}|A_t\rangle = Np(A_t)=N-N\left(1-\frac{m}{N}\right)^{A_t}\left(1-\lambda(h)\right). 
\end{equation}  
We introduce the effective branching parameter $\meff(A_t)$ to satisfy
Eq.~\eqref{eqBPcondexp}, i.e.,  
\begin{equation}
  \langle A_{t+1}|A_t\rangle =\meff(A_t) A_t + \lambda(h)N, 
\end{equation}
such that
\begin{equation}\label{eqResultAnaMeff}
  \meff(A_t,N,m,h)=
  \left(\frac{N}{A_t}\right)\left(1-\left(1-\frac{m}{N}\right)^{A_t}\right)\left(1-\lambda(h)\right).
\end{equation}
%

We can further compute the network activation rate $A=\langle A_t\rangle$. For
this, we recall $\langle A_{t+1}|A_t \rangle = Np(A_t)$ and assume stationary
activity, where the law of total expectation yields $A=\langle
A_t\rangle=\langle\langle A_{t+1}|A_t \rangle\rangle$. Combined with the
approximation $\langle(1-m/N)^{A_t}\rangle\approx (1-m/N)^{\langle A_t\rangle}$,
Eq.~\eqref{eqResultAnaCond} results in the mean-field approximation 
\begin{equation}\label{eqResultAnaMFapproximation}
  A = N-N\left(1-\frac{m}{N}\right)^{A}\left(1-\lambda(h)\right).
\end{equation}
To solve Eq.~\eqref{eqResultAnaMFapproximation} for the network rate $A$, we
first rewrite it as
\begin{align}
  (N-A)\ln&\left(1-\frac{m}{N}\right)e^{(N-A)\ln(1-\frac{m}{N})}\nonumber\\
  &=\ln\left(1-\frac{m}{N}\right)N \left(1-\frac{m}{N} \right)^N(1-\lambda(h)),
\end{align}
and make use of the Lambert-W function, defined by $W(z) e^{W(z)} =
z$~\cite{corless1996, valluri2000}, to obtain
\begin{equation}
  A(N,m,h)= N -
  \frac{W\left[\ln(1-\frac{m}{N})N(1-\frac{m}{N})^N(1-\lambda(h))\right]}{\ln(1-\frac{m}{N})}.
\end{equation}
In the limit $N\to\infty$, we can replace \mbox{$\ln(1-\frac{m}{N})N\to-m$} and
$(1-\frac{m}{N})^N\to e^{-m}$, to obtain for the activation rate per node
$a=A/N$
\begin{equation}\label{eqResultAnaResponse}
  a(m,h) \xrightarrow[N \to \infty]{}  1+\frac{W[-me^{-m}(1-\lambda(h))]}{m}.
\end{equation}
\begin{figure}
  \includegraphics{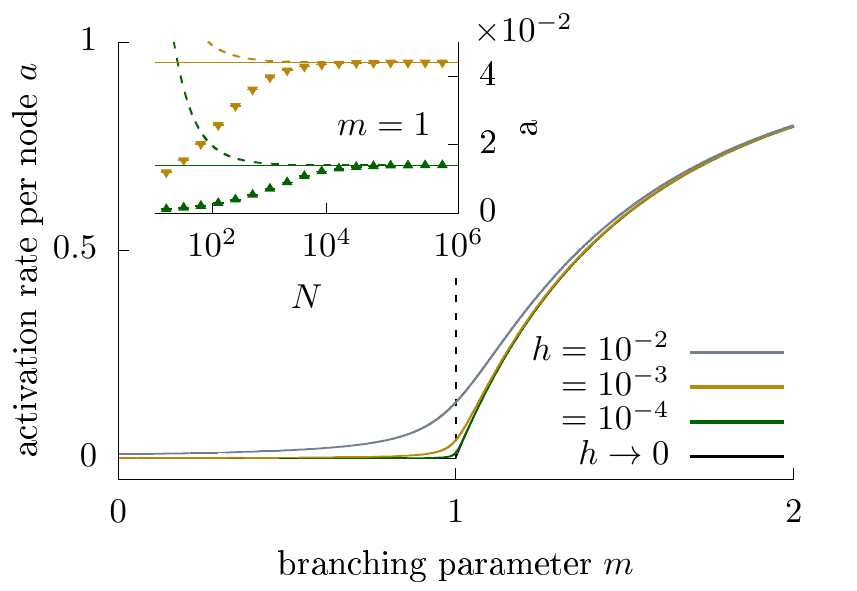}
  \caption{%
    Analytic solution for the activation rate per node ($a$) as a function of
    the branching parameter ($m$) in a mean-field branching network
    ($N\to\infty$) for different external input rates ($h$, encoded by color).
    For $h\to0$, the branching network undergoes a critical non-equilibrium
    phase transition from an absorbing phase ($a=0$) to an active phase
    ($a>0$). \textit{Inset}: $a$ as a function of network size $N$ at the
    critical point ($m=1$), comparing the analytic solution (dashed lines) and
    asymptotic limit (solid lines) with simulation results (data points) for
    exemplary external input rates (color as in main plot).  
    \label{figFssDrivenActivity}
  }
\end{figure}

This asymptotic solution characterizes the phase diagram of the mean-field
branching network (Fig.~\ref{figFssDrivenActivity}): for $h\to 0$ a critical
point ($m=1$) separates an absorbing phase ($m<1$, $a_t=0$) from an active phase
($m>1$, $a_t\neq0$). Formally, for $h\neq 0$ there is no critical
non-equilibrium phase transition~\cite{henkel2008}. For a specific connectivity,
behavior of the driven system was shown to demonstrate
quasi-criticality~\cite{williams-garcia2014}. We will refer to $m=1$ as
critical-like dynamics, because the external input rates we consider here are very
small, and therefore the dynamics is very similar to true critical dynamics on
finite networks.

Our numerical results verify that in the limit $N\to\infty$ the rate per node
converges to the analytic solution (Fig.~\ref{figFssDrivenActivity}, inset). The
difference for small system sizes can be explained by the presence of a
temporally absorbing state ($A_t=0$), where the system stays silent until the
new external input has arrived. This state is not captured by the mean-field
approximation, Eq.~\eqref{eqResultAnaMFapproximation}, which assumes stationary activity, and therefore overestimates
$a$ for small $N$. 

The phase diagram of the mean-field branching network is qualitatively different
from a branching process. For the branching process, the critical point
($\hat{m}=1$) separates the subcritical (stable) phase ($\hat{m}<1$) from the
supercritical (divergent) phase ($\hat{m}>1$). This is already a strong
indication for the potential emergent bias when approximating network activity
in a branching network by a branching process. 
In order to compare with the branching-process approximation, we nonetheless
use the notion of subcritical-like ($m<1$), critical-like ($m=1$), and
supercritical-like ($m>1$) spreading dynamics for the branching network.

\subsection{Analytic derivations for the result of branching-process
approximations in driven branching networks}
\label{secResultAnaPrediction}

In the following, we make use of the three estimators introduced in
Sec.~\ref{secModel} to derive the results of branching-process approximations
to an asymptotically large branching network, based on our analytical results
on the effect of coalescence.

\begin{figure}[t]
  \includegraphics{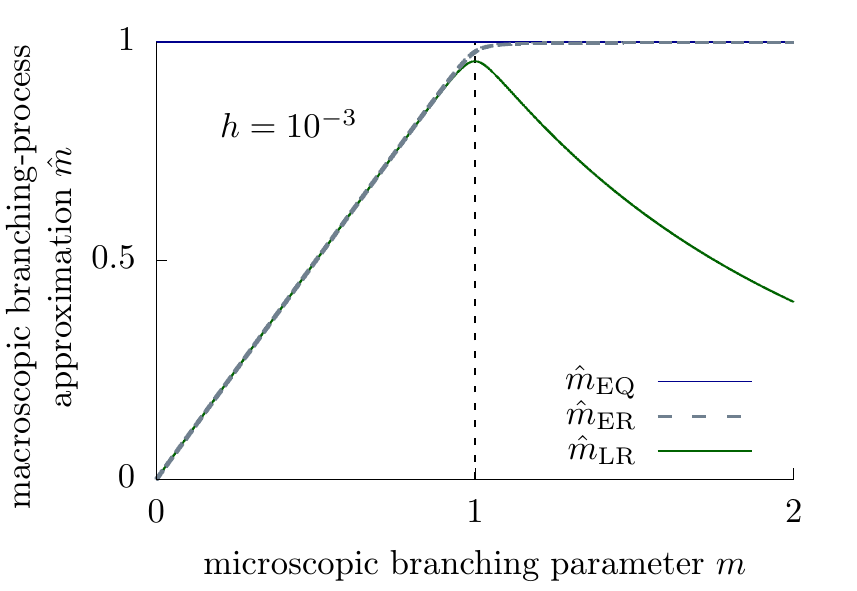}
  \caption{%
    Prediction of the asymptotic ($N\to\infty$) macroscopic branching parameter estimates
    $\hat{m}$ 
    as a function of the microscopic
    (model) branching parameter $m$ of a driven branching network. The drive is chosen relatively large
    ($h=10^{-3}$) to make the effects around $m=1$ visible.
    \label{figResultPredictionBias}
   }
\end{figure}

The simplest case is the branching process approximation through expected
network rate, $\mER$. Here, we can simply insert the mean-field solution of
the network rate, Eq.~\eqref{eqResultAnaResponse}, and obtain
\begin{equation}
  \mER(m,h) = 1 - \frac{mh\Delta t}{m+{W[-me^{-m}(1-\lambda(h))]}}.
\end{equation}
If the external input rate $h$ is known, which is usually not the case, this
estimator is not as biased as one would naively expect
(Fig.~\ref{figResultPredictionBias}, dashed gray line). For $m<1$, the
estimator $\mER$ works well. However, it starts to get biased around $m=1$ and
for $m>1$ saturates to its upper bound $\mER\to1$. This bias is due to the fact
that in the branching process, stationary activity can only be realized for
subcritical dynamics ($m<1$). Thus in general, we expect a good approximation
of a branching network by a branching process only in the subcritical regime.

Next, we calculate the asymptotic result of the branching-process approximation
through the expected quotient of subsequent network-activity, $\mEQ$. This
estimator has been frequently applied to process neural data. For our
derivation, we assume a driven network with $Nh\gg1$, so that the network
activity is practically always nonzero, i.e., $A_t>0$. This is similar to the
case of increasing bin size when processing neural data~\cite{priesemann2014}.
Assuming an effective branching process with $A_t>0$ for all $t$, we can
approximate Eq.~\eqref{eqModelEstimateBP} as $\mEQ \approx
1+Nh\left(\langle\frac{1}{A_t}\rangle - \frac{1}{\langle A_t\rangle} \right)$
(see Appendix~\ref{appApproxEQ}). Away from the critical point, one can expect
that fluctuations around $A_t$ vanish in the limit of large system sizes such
that $\langle\frac{N}{A_t}\rangle\approx \frac{N}{\langle A_t\rangle}$ and 
%
%
%
\begin{equation}\label{eqResultAnaPredmEQ}
  \lim_{N\to\infty}\mEQ(m,h) = 1 \qquad \forall\, m\,\,\text{and}\,\, h>0.
\end{equation}
By Jensen's inequality for the convex function $1/x$ we find
$\langle\frac{N}{A_t}\rangle> \frac{N}{\langle A_t\rangle}$ such that $\mEQ$
should approach its limit from above. Therefore, in the limit $N\to\infty$
(fixed $h>0$) any stationary activity in the driven regime is interpreted as a
persistent internal spread, i.e., $\mEQ$ always infers $m=1$
(Fig.~\ref{figResultPredictionBias}, solid blue line). This is because  the
definition of $\mEQ$ implicitly assumes a STS or equivalently $h \to 0$, and
under this assumption only $m=1$ would produce stationary activity on
expectation. Hence it is a correct estimator, as long as there is only internal
activation, but fails as soon as novel input $h$ is applied while the network
is still active ($A_t>0$). The estimator thus does what it is supposed to do,
but does not help us to quantify the amount of internal activation in any
driven regime like the living brain.

\begin{figure}[t]
  \includegraphics{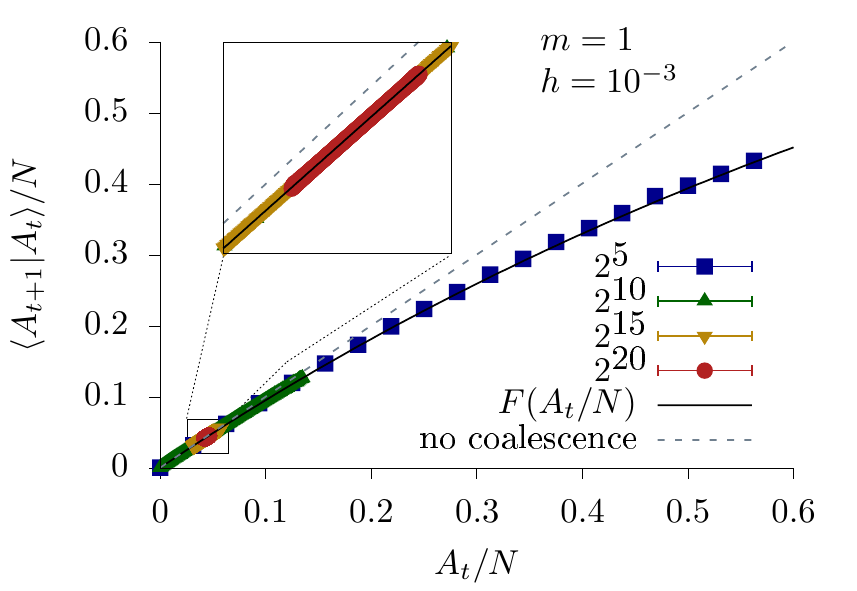}
  \caption{%
    Universal scaling function describes effective spreading of network
    activity for sufficiently large system sizes (here $N\geq2^5$, $m=1$ and
    $h=10^{-3}$).  Rescaling the conditional expectation value as $\langle
    A_{t+1} | A_t\rangle/N$ (data points) leads to a data collapse onto the
    universal scaling function $F(A_t/N)$ (solid line) defined in
    Eq.~\eqref{eqScaling}. Due to the non-linear character of $F(A_t/N)$, a
    linear-regression estimate results in a slightly different slope from the
    case without coalescence (dashed line). 
    \label{figScalingFunction}
    }
\end{figure}

The third estimator, $\mLR$, is based on a branching-process approximation
through linear regression. It does not rely on knowledge of $h$ and does not
require any STS or specific regime for $A_t$. This estimator returns reliable
results in the subcritical regime. In the vicinity of $m=1$, we show that its
asymptotic result can be fully attributed to non-vanishing coalescence effects.
In detail, the asymptotic estimate can be calculated from the conditional
expectation value $\langle A_{t+1}|A_t\rangle$. Normalizing
Eq.~\eqref{eqResultAnaCond}, we find a system-size independent scaling function
$ F(A_t/N)$ for the activity per node $a_t=A_t/N$:
\begin{align}
  \langle A_{t+1}|A_t\rangle/N 
  &=  1-\left(1-\frac{m}{N}\right)^{A_t}(1-\lambda(h))\nonumber\\
  &=  1-\left(1-\frac{m}{N}\right)^{N A_t/N}(1-\lambda(h))\nonumber\\
  &\simeq 1-e^{-mA_t/N}(1-\lambda(h))\nonumber\\
  &= 1-e^{-(h\Delta t+mA_t/N)} = F(A_t/N).\label{eqScaling}
\end{align}
Indeed, numerical results of the normalized conditional expectation value
covering system sizes from $2^5$ up to $2^{20}$ all collapse onto this universal scaling
function (Fig.~\ref{figScalingFunction}, $m=1$ and $h=10^{-3}$). With increasing
system size, the variance of the average activity decreases and the numerical
results localize along the scaling function, justifying the mean-field
assumptions above.  

\begin{figure*}
  \begin{center}
    \begin{tikzpicture}
      \def\x{9.00cm}
      \def\y{5.5cm}
      \def\dx{-4.0cm}
      \def\dy{ 2.6cm}
      \node (a) at (-\x, 0cm) {\includegraphics{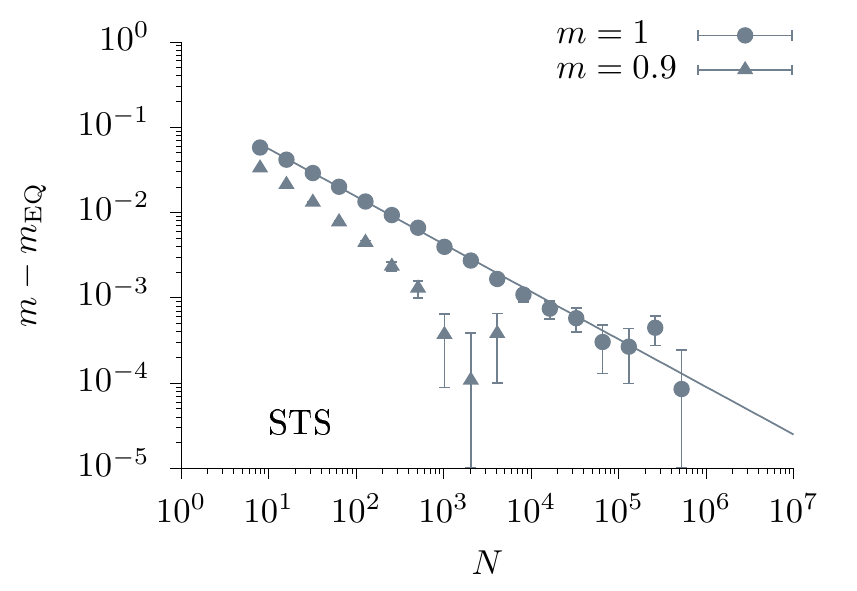}};
      \node (b) at (0cm, 0cm) {\includegraphics{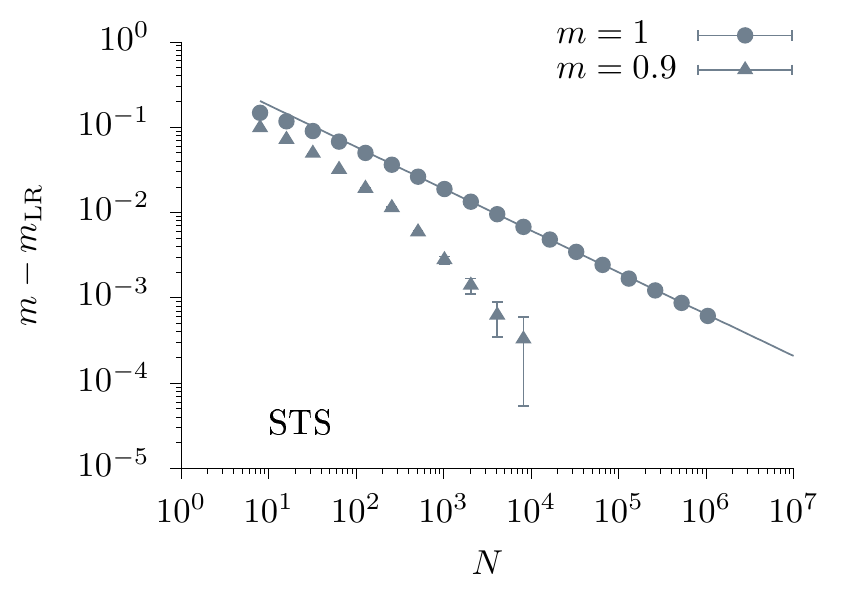}};
      \node[at={($(a)+(\dx,\dy)$)}]{$\textbf{A}$};
      \node[at={($(b)+(\dx,\dy)$)}]{$\textbf{B}$};
    \end{tikzpicture}
  \end{center}
  \caption{%
    Finite-size scaling in the separation-of-timescale regime for two standard
    estimators of the macroscopic branching parameter, \textbf{A} the
    expected-quotient estimator $\mEQ$, and \textbf{B} the  linear-regression estimator    $\mLR$. Both estimators converge to the microscopic branching parameter for $m\leq 1$. In the
    absorbing phase ($m=0.9$), the macroscopic estimate quickly converges
    towards the microscopic model parameter. For critical dynamics ($m=1$) the
    macroscopic estimates converge towards the microscopic parameter as a
    power law $1-\hat{m}_x\sim N^{-\alpha}$. 
    \label{figFssSts}
    }
\end{figure*}

\begin{figure*}
  \begin{center}
    \begin{tikzpicture}
      \def\x{9.00cm}
      \def\y{5.5cm}
      \def\dx{-4.0cm}
      \def\dy{ 2.6cm}
      \node (a) at (-\x, 0cm) {\includegraphics{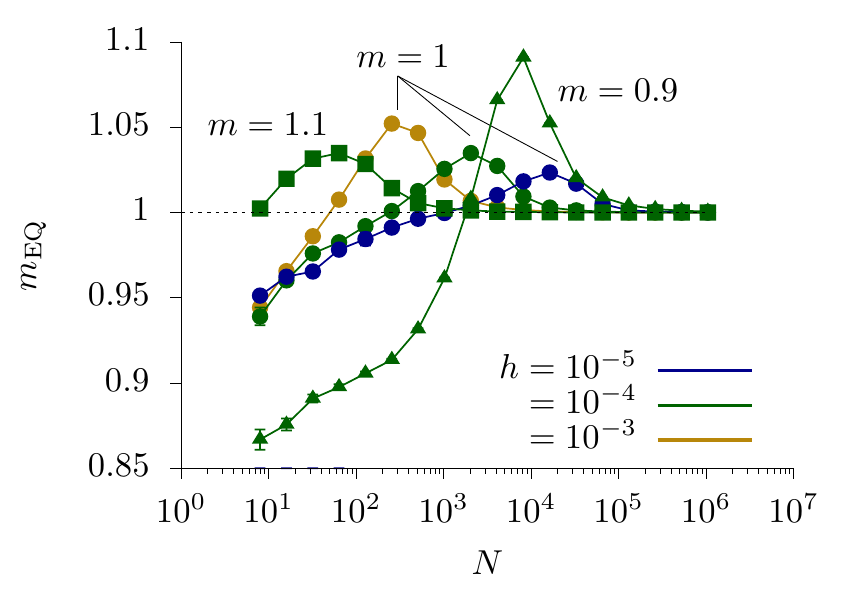}};
      \node (b) at (0cm, 0cm) {\includegraphics{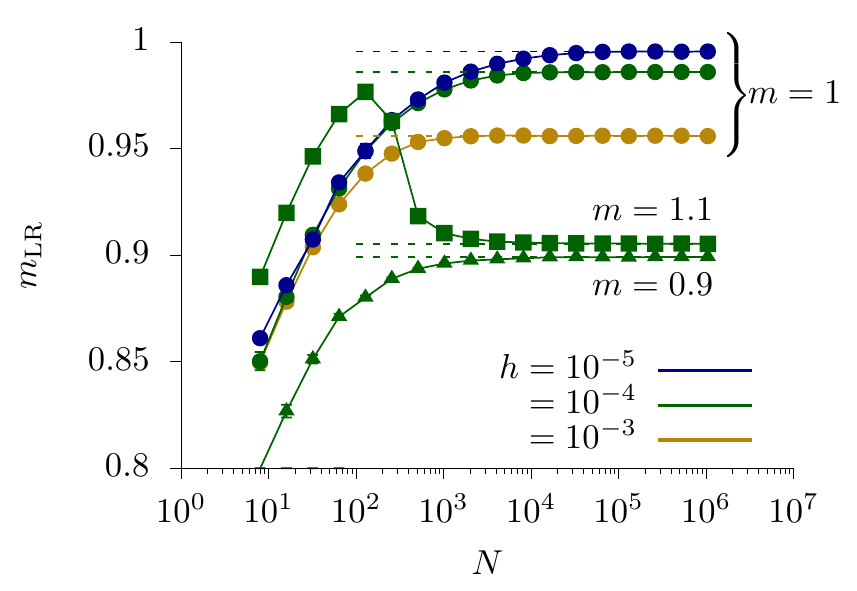}};
      \node[at={($(a)+(\dx,\dy)$)}]{$\textbf{A}$};
      \node[at={($(b)+(\dx,\dy)$)}]{$\textbf{B}$};
    \end{tikzpicture}
  \end{center}
  \caption{%
    Finite-size scaling in the driven regime. The two
    estimators of the macroscopic branching parameter, \textbf{A} the
    expected-quotient estimator $\mEQ$, and \textbf{B} the  linear-regression estimator   $\mLR$ both fail to measure the microscopic
    branching parameter due to coalescence effects (full lines are a guide to the eye) Dashed lines are the
    calculated infinite-size limits of the respective estimates (see text). 
    \label{figFssDriven}
    }
\end{figure*}

From the curvature of the scaling function we can derive the asymptotic result
of the linear-regression estimator $\mLR$. The linear regression assumes a
linear shape of the conditional expectation value (independence of $A_{t+1}$ on
$A_t$). As a consequence, $\mLR$ locally fits a straight line to the scaling
function with diminished slope (Fig.~\ref{figScalingFunction}). The slope
of this line depends on the rate per node. For non-zero rate, it will deviate
from the ideal case without coalescence (Fig.~\ref{figScalingFunction}, dashed
line) and thereby the estimate will differ from the model parameter. Because the
variance of the rate per node decreases with increasing system size, we can
calculate the asymptotic estimate as the derivative of Eq.~\eqref{eqScaling} at
the average rate, Eq.~\eqref{eqResultAnaResponse}, i.e.,
\begin{align}\label{eqResultAnaPredmLR}
  \mLR(m,h)
  &=\left.\frac{d}{d A_t/N} F(A_t/N) \right|_{A(m,h)/N}\nonumber\\
  &= m e^{-m\, a(m,h)}(1-\lambda(h))\\
  &= m e^{-m-h\Delta t}e^{-W[-m e^{-m-h\Delta t}]}.\nonumber
\end{align}
The asymptotic estimate $\mLR$ is thus biased in the vicinity of the critical
point and the active phase (Fig.~\ref{figResultPredictionBias}, solid green line
and Fig.~\ref{figFssDriven}, dashed lines). Importantly, the asymptotic bias
vanishes for $h\to0$ only within the absorbing phase ($m\leq 1$). 


To summarize, in a driven regime all estimators of the microscopic dynamics are
biased close to critical-like settings. Some are biased for the whole parameter
range and reflect only the presence of drive (such as $\mEQ$). Others are
deviating slightly from the true values for $m < 1$, but are strongly
biased for  $m \gtrapprox 1$, (Fig.~\ref{figResultPredictionBias}). The reason
for the asymptotic bias is the coalescence in branching networks
(Sec.~\ref{secResultAnaCoalescence}). In the following subsections, we will
numerically verify our analytical predictions on the estimation bias with
finite-size scaling analyses of branching networks, both for the STS
(Sec.~\ref{secResultFssSTS}) and the driven (Sec.~\ref{secResultFssDriven})
regime.

\subsection{Finite-size scaling analysis in separation-of-timescale (STS) regime reveals no asymptotic bias in branching-process approximation}
\label{secResultFssSTS}
In the STS regime, activity (also called an avalanche) is initiated at a random
node and evolves without any external input until the end. Formally, this
corresponds to the limit of vanishing external input rate $h\to0$. In numerical
implementations one can skip periods of zero activity: Directly after activity
has ceased, a new random node is initiated at $t+1$. In this regime the
internal network dynamics depends only on the model parameter $m$. The
finite-size scaling limit is well defined for $N\to\infty$ by keeping $m$
fixed. 
%

We focus on the two estimators $\mEQ$ and $\mLR$. The third one, $\mER$,
requires stationary activity that is not present in the STS regime. In principle
both estimators are suitable for the formal STS regime, but the numerical
implementation (skipping periods of zero activity) requires a little attention.
The expected-quotient estimator $\mEQ$ is not affected by skipping periods with
zero activity, because it only considers time steps with $A_t>0$. The
linear-regression estimate $\mLR$ is, however, strongly affected by the choice
of skipping periods of zero activity, because periods of zero activity contribute
to the estimated external input rate relevant for the linear regression. We thus
need to modify the linear-regression estimator for the STS regime by imposing a
zero expectation of network activity $\langle A_t\rangle=0$. This enters both
the covariance and variance and we obtain
\begin{equation}
  \mLR^\mathrm{STS} = \frac{\langle A_{t+1} A_t\rangle}{\langle
  A_t^2\rangle}.
\end{equation}
Keep in mind that this is only required because of the artificial implementation
of the STS regime ($h\to 0$) as non-stationary activity. In addition, the
implicit assumption of vanishing network rate $\langle A_t\rangle=0$ confines
our discussion to the  absorbing phase $m\leq 1$.

In the STS regime, one may expect that the branching-process approximation from
network activity is not biased in the limit $N\to\infty$. This is because
small avalanches and equivalently small $A_t$ occur statistically more often
than large avalanches, even for critical spreading dynamics. More precisely,
the same small-avalanche regime of the characteristic avalanche-size
distribution remains dominant with increasing system size. At the same time,
for annealed disorder the number of potential connections increase with system
size, such that instances of internal coalescence become less probable.
Together, we expect that in the STS regime macroscopic estimates reflect
microscopic dynamics in the limit $N\to\infty$.

Indeed, the bias in the estimates  of $\mEQ$ and
$\mLR^\mathrm{STS}$ decreases rapidly with increasing system size $N$ (Fig.~\ref{figFssSts}).
In the absorbing phase ($m=0.9$), the bias diminishes rapidly, and
is below $0.001 \%$ for $N>10^4$.
For critical dynamics ($m=1$), the bias decreases as a power-law $1-\hat{m}_\mathrm{x}\propto N^{\alpha_\mathrm{x}}$.
Least-square fits yield $\alpha_\mathrm{EQ}=0.558(7)$, for $N>32$ with
$\chi^2\approx1.4$, and $\alpha_\mathrm{LR}=0.490(2)$, for $N>512$ with
$\chi^2\approx 0.3$. Upon changing the fit range, however, estimates vary
outside of their statistical errors, compatible with an overall scaling of the
form $1-\hat{m}\propto N^{-1/2}$.

The STS regime allows to directly investigate internal coalescence because
external coalescence is excluded. Our results that the microscopic model
parameter ($m\leq1$) can be correctly estimated from the network activity either
via $\mEQ$ or $\mLR$ in the limit $N\to\infty$ shows that the asymptotic effect
of internal coalescence vanishes in the absorbing phase ($m\leq1$).

\subsection{Finite-size scaling analysis in (Poisson) driven regime reveals
asymptotic bias in branching-process approximations}
\label{secResultFssDriven}
We now consider the driven regime of a network subject to homogeneous Poisson
input. Considering additional stochastic external input, we need to specify how
the infinite-size limit ($N\to\infty$) is approached: Here, we choose to fix the
microscopic model parameter $m$ as well as the average external input rate per
node $h$. This assumes that the external input rate scales with system size.  We
focus on the two estimators $\mEQ$ and $\mLR$, because they do not rely on
knowledge of $h$. 

We first show that in the driven regime the commonly employed expected-quotient
estimator $\mEQ$ will always indicate critical-like dynamics ($m=1$) for large
system sizes as predicted by Eq.~\eqref{eqResultAnaPredmLR}
(Fig.~\ref{figFssDriven}~A). If the dynamics is indeed critical like, $\mEQ$
first underestimates the microscopic dynamics for small $N$ ($\mEQ<1$),
overestimates them for intermediate $N$ ($\mEQ>1$), and finally converges to the
true microscopic dynamics for $N\to\infty$. The regime of overestimation shifts
to larger systems sizes and decreases its amplitude as the input rate decreases.
While this would seem reasonable for $m=1$, the estimator $\mEQ$ fully fails for
subcritical-like ($m=0.9$) and supercritical-like ($m=1.1$) dynamics. 
Independently of the true $m$, being it smaller, larger or equal to unity, the
estimator returns $\mEQ<1$ for small $N$, a local maximum with $\mEQ>1$ for
intermediate $N$, and  eventually converges towards $\mEQ\to1$ in the limit
$N\to\infty$.  This is clearly not the microscopic dynamics. 
Strikingly, $\mEQ$ would thereby predict critical-like dynamics for
sufficiently large networks $N>10^5$ even though the microscopic dynamics are
clearly not critical like. 

The system-size dependence of $\mEQ$, showing a maximum ($\mEQ>1$) and
converging towards  $\mEQ\to1$ in the limit $N\to\infty$, is similar to results
obtained when changing the bin size for neural spike
recordings~\cite{priesemann2014}. The initial increase and maximum can therefore
be explained as extended periods of activity separated by a decreasing number of
time steps with zero activity.  The eventual convergence towards unity can then
be explained by the resulting stationary activity and the absence of zero
activity due to the increasing amount of network input $hN$
(Sec.~\ref{secResultAnaPrediction}).

Next, we show that in the driven regime the linear-regression estimator $\mLR$
underestimates microscopic dynamics (Fig.~\ref{figFssDriven}~B). While this
estimator is specifically constructed to infer $m$ from driven systems, it does
not consider coalescence. Coalescence leads to a bias even in the infinite-size
limit as predicted by Eq.~\eqref{eqResultAnaPredmLR} and verified by our
numerical results (Fig.~\ref{figFssDriven}~B, dashed lines). The system size
above which $\mLR$ saturates, corresponds to the system size above which also
the rate per node saturates (Fig.~\ref{figFssDrivenActivity}, inset). 

We want to point out two things. First, the asymptotic bias of $\mLR$ depends on
$h$, demonstrated here only for critical-like dynamics ($m=1$), where the effect
of $h$ is strongest. Second, in the limit $h\to 0$ the asymptotic bias of $\mLR$
only vanishes for $m\leq 1$. This means that for supercritical-like dynamics
($m>1$) the asymptotic estimator $\mLR$ remains biased even for $h\to0$ and
indicates subcriticality $\mLR<1$ (Fig.~\ref{figResultPredictionBias}).
However, in the supercritical-like regime, the activity per node can obviously
become quite high, even with little input.


\subsection{Coalescence can be captured by a non-linear estimator}
\label{secResultFssNonLin}
\begin{figure}
  \includegraphics{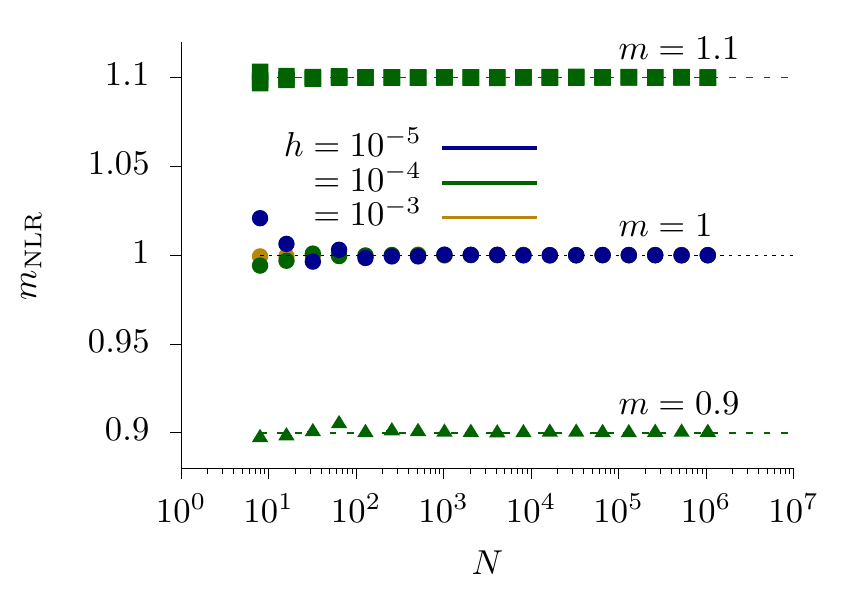}
  \caption{%
    The novel non-linear estimator $\mNLR$ correctly infers microscopic model
    parameter $m$ from the macroscopic dynamics in all-to-all connected
    branching networks for supercritical-like ($m=1.1$), critical-like ($m=1$),
    and subcritical-like ($m=0.9$) dynamics.
    \label{figFssNLR}
  }
\end{figure}
We can use our analytic results to obtain the model parameter $m$ without bias
from the macroscopic network activity by directly fitting the non-linear
function, Eq.~\eqref{eqScaling}, to the data. This defines our non-linear regression
estimator $\mNLR$ as a fit to the non-linear scaling function
\begin{equation}
  \langle A_{t+1}|A_t\rangle/N = 1-e^{-(h\Delta t + \mNLR A_t/N)}.
\end{equation}
We implemented this as a python curve fit. For our numerical data this
non-linear approach correctly infers the microscopic model parameter from the
macroscopic network activity for almost all system sizes (Fig.~\ref{figFssNLR}). Of
course, this relies on a universal scaling function for the conditional
expectation value $\langle A_{t+1} |A_t\rangle=NF(A_t/N)$, here derived for an
annealed disorder average. A similar scaling function has been derived within a
mean-field approximation for quenched disorder average over Erd\H{o}s-R{\'e}nyi
networks with average degree $K$~\cite{kinouchi2006}:
\begin{equation}\label{eqKC}
  F(A_t/N)=1-\left(1-\frac{m A_t}{N K}\right)^K(1-\lambda(h)),
\end{equation}
where we neglected the refractory period to compare with our results. Expanding
Eq.~\eqref{eqKC} agrees to leading order with the expansion of
Eq.~\eqref{eqScaling}. 

If analytic results are not available in practice, one can try to estimate the
universal scaling function from small system sizes. Fortunately, small system
sizes have a larger variance in the average activity $a$ (cf.
Fig.~\ref{figScalingFunction}). Simulating several small system sizes would
thereby allow one to rescale the axes until one obtains a collapse of data
points. The resulting data collapse then yields the universal scaling function.

\section{Discussion}
\label{secDiscussion}
To summarize, we have shown that due to coalescence (the simultaneous
activation of the same node from multiple sources) in a branching network, the
approximation of network activity by a branching process can be biased. As the
branching-process approximation of network activity is the basis for several
linear estimators of spreading dynamics on networks, these estimators can be
consequently biased as well. We verified this bias for an estimator based on
the expected quotient of subsequent activities ($\mEQ$) and an estimator based
on linear regression ($\mLR$). In the separation-of-timescale (STS) regime,
which we argued is only well-defined in the absorbing phase ($m\leq1$), the
bias vanishes for $N\to\infty$.
In the driven regime of non-vanishing input rate, there always remains an
asymptotic bias for $N\to\infty$. We showed how to analytically compute the
asymptotically remaining bias in the driven regime and verified it by a
finite-size scaling analysis of simulation results. 

When it comes to approximating real data with a macroscopic branching process,
the potential bias (assuming that there is a static interaction network) can
be evaluated on a case-to-case basis. For example, cortical neural network
dynamics have been estimated to be slightly subcritical with
$\hat{m}\approx0.98$~\cite{wilting2018,wilting2019}. This indicates that $2\%$
of the activity is generated by external input. For a neural firing rate per
neuron of $\mathcal{O}(\SI{e-3}{1/ms})$ one thus expects an external input rate
$h=\mathcal{O}(\SI{e-5}{1/ms})$. This assumes that activity propagates with
time steps of $\Delta t=\SI{1}{ms}$.
These estimates are consistent with numerical predictions for cortical network
dynamics suggesting $h\Delta t=\mathcal{O}\left(10^{-4}\right)$ or
lower~\cite{zierenberg2018}. In this case, our results for the
linear-regression estimator would predict a bias $m-\mLR =
\mathcal{O}\left(10^{-3}\right)$.  This bias is an order of magnitude smaller
than the typical observed values for cortical brain networks \textit{in vivo}
($\hat m \approx 0.98$)~\cite{wilting2018}. The largest observed values though
reach $\hat m \approx 0.994$~\cite{wilting2019}, suggesting that certain
neural networks approach the critical point almost as close as possible, given
coalescence.

It is a priori unclear whether the microscopic model parameter $m$ or the
macroscopic parameter $\hat{m}$ is the correct description of the dynamics. We
expect that the most reliable estimator of macroscopic dynamics is the
linear-regression estimator $\mLR$, because it explicitly considers the external
input rate present in many practical situations~\cite{heyde1972, wei1990,
wilting2018}.  Whether inferring the microscopic or macroscopic branching
parameter is ``correct'', however, depends on which question is asked: On the
one hand, if we want to infer the microscopic dynamics to understand microscopic
processes, we seek $m$. For example, to characterize the impact of a single
spike on the amount of subsequent spike initiations~\cite{london2010}, or to
predict probable routes of disease spreading in complex
networks~\cite{brockmann2013}. On the other hand, if we want to describe the
time evolution of the collective network activity, we are interested in the
macroscopic $\hat{m}$. For example, to characterize intrinsic timescales of
cortical areas~\cite{murray2014} or to estimate whether or not a disease has
epidemic character~\cite{farrington2003,wilting2018}. Another option could be
that reality implements coalescence-compensating
mechanisms~\cite{zierenbergShort}, thereby interpolating between both cases. The
interpretation of ``correct'' branching parameter thus depends on whether we
want to know the microscopic or the macroscopic dynamics of a particular system.

If interested in an unbiased estimator of the microscopic branching parameter
from the network activity, we provide a non-linear estimator that explicitly
takes coalescence into account and is thereby not biased. The non-linear
estimator is derived from the analytic solution of the conditional expectation
value of subsequent network activities $\langle A_{t+1} | A_t\rangle$. This
conditional expectation value has a universal scaling function for annealed
disorder (derived here) and quenched disorder over random Erd\H{o}s-R{\'e}nyi
networks (derived in Ref.~\cite{kinouchi2006}). If the scaling function is not
known analytically, we propose that it can be obtained by inducing a data
collapse for the conditional expectation value measured for small system sizes
with high numerical precision (cf.~Fig.\ref{figScalingFunction}). While this
approach is directly applicable to models, it cannot be applied trivially to
experimental data. For one, the scaling function would need to be deduced from
a representative model. In addition, our current results require fully-sampled
network activity. However, advances in recording techniques, e.g., optogenetic
imaging of neural activity~\cite{nagel2002, boyden2005, kim2017} or high report
rates for measles in Germany~\cite{wilting2018}, may enable to construct
non-linear estimators applicable to experimental data even in large systems.

Due to coalescence, the non-equilibrium phase transition in branching networks
differs from that in branching processes. On the one hand, for branching
networks without external input, the critical point separates an absorbing
($A=0$) from an active ($A>0$) phase, a critical phase transition in the
universality class of directed percolation~\cite{hinrichsen2000,henkel2008}. In
fact, the branching network defined in this work is equivalent to mean-field
directed bond percolation. Here, the order parameter is the network activity.
On the other hand, for branching processes the critical point separates a
subcritical (zero probability for infinite avalanche or activity) from a supercritical
(non-zero probability for infinite avalanche) phase~\cite{harris1963}. Here,
the order parameter is the probability for infinite avalanches. However, the
expected population activity for a subcritical branching process is indeed zero.
Hence, branching network and branching process share universal features in the
absorbing or subcritical phase, while their activities vary substantially in
their active or supercritical phase.

We considered in our study homogeneous external input rates per node. This is
clearly a leading-order approximation. In the context of neural networks, a
homogeneous input rate per node can be motivated by network-wide input that
cortical areas receive. In the context of infectious diseases, a homogeneous
external input rate corresponds to a homogeneous external infection rate
throughout the environment. It is natural to expect that input rates are in fact
more heterogeneous. In the context of neural networks, already the functional
wiring of cortical layers induces heterogeneity, e.g., see
Ref.~\cite{felleman1991,harris2015}. In the context of infectious diseases, it
seems natural that external infection rates depend on local environmental
variables, e.g., see Ref.~\cite{real2007}. We expect that heterogeneous input
rates will contribute an additional source of bias in the branching-process
approximation of spreading dynamics. 

We focused in our study on the case of annealed disorder, mathematically
equivalent to an all-to-all connected network. This mean-field assumption turns
out to be a good leading-order approximation for many complex
networks~\cite{gomez2010}. Our annealed disorder also covers memory-less
temporal networks~\cite{moinet2018}. However, we have also shown that several of
our analytical results agree to leading-order with those obtained for quenched
disorder of random networks~\cite{kinouchi2006}. We hence expect that the
majority of our conclusions is valid to leading order for general random
networks, but also expect that details depend on the considered network
topology. For example, the vanishing bias in the STS regime requires a large
number of connections per node in the limit $N\to\infty$. This would be
guarantied if the average degree $K$ would be coupled to the system size, i.e.,
$K/N$ constant. However, reality may be quite different. In the scope of
neuroscience, one expects sparsely connected networks with
$K/N\to0$~\cite{brunel2000}. Also in the scope of infectious diseases in human
contact networks, one expects a finite number of interaction
links~\cite{salathe2010}. Moreover, there is an increasing number of studies
that identify aspects of heterogeneous network topology that affect collective
network dynamics in general~\cite{watts1998,kitsak2010, ronellenfitsch2018}, or
specifically collective dynamics in neural
networks~\cite{moretti2013,orlandi2013,levina2014theoretical, effenberger2015,
mastrogiuseppe2018,bondanelli2018} and for infectious
diseases~\cite{newman2002,shirley2005,hindes2016}. We expect that the
branching-process bias remains a general leading-order effect in heterogeneous
network topologies. 

Our results are only applicable to fully-sampled systems. However, typically experimental
measurements only have access to a small fraction of the system, resulting in
subsampled data. Subsampling is a common problem in
neuroscience~\cite{priesemann2009,ribeiro2014, levina2017, wilting2018,
wilting2019}, epidemiology~\cite{papoz1996, cormack1999, sethi1999, chao2001,
tilling2001}, and networks in general~\cite{stumpf2005}. In the case of
subsampled network activity, there is an unbiased way to estimate the effective
branching parameter $\meff$ by extending the linear-regression estimator to
multiple regressions~\cite{wilting2018}. We leave it for future work to
expand our results on convergence effects to the subsampled regime.

\begin{acknowledgments}
JZ would like to thank Peter Sollich, Peter Grassberger and Malte Henkel for
  fruitful discussions. JZ and VP received financial support from the German
  Ministry of Education and Research (BMBF) via the Bernstein Center for
  Computational Neuroscience (BCCN) G{\"o}ttingen under Grant No.~01GQ1005B. JW
  was financially supported by Gertrud-Reemtsma-Stiftung.  AL received funding
  from a Sofja Kovalevskaja Award from the Alexander von Humboldt Foundation,
  endowed by the Federal Ministry of Education and Research.
\end{acknowledgments}

\begin{appendix}
  \section{Evaluation of branching-process approximation through expected
  quotient ($\mEQ$) in the driven regime}
  \label{appApproxEQ}
  In the driven regime, we assume that all \mbox{$A_t>0$} for \mbox{$Nh\gg0$}.
  We can then evaluate Eq.~\eqref{eqModelEstimateBP}, using Bayes' rule, \mbox{$P[A_{t+1}|A_t] = P[A_{t+1},A_t]P[A_t]$},
  \begin{align*}
    \left\langle\frac{A_{t+1}}{A_t}\right\rangle 
    &=\sum_{A_t}\sum_{A_{t+1}}P[A_t, A_{t+1}]\frac{A_{t+1}}{A_t}\\
    &=\sum_{A_t}\frac{1}{A_t}P[A_t]\sum_{A_{t+1}}\frac{P[A_t, A_{t+1}]}{P[A_t]}A_{t+1}\\
    &=\sum_{A_t}\frac{1}{A_t}P[A_t]\sum_{A_{t+1}}P[A_{t+1}|A_t] A_{t+1}\\
    &=\sum_{A_t}\frac{\langle A_{t+1} | A_t\rangle}{A_t}P[A_t].
  \end{align*}
  Assuming an effective branching process, we insert $\langle A_{t+1} |
  A_t\rangle = \hat{m}A_t + hN$ with $\hat{m}=1-hN/\langle A_t\rangle$ and
  obtain
  \begin{align*}
    \left\langle\frac{A_{t+1}}{A_t}\right\rangle 
    &\approx\sum_{A_t}\frac{\hat{m}A_t + hN}{A_t}P[A_t]\\
    &\approx\hat{m} + hN\left\langle \frac{1}{A_t}\right\rangle\\
    &\approx 1 + hN\left(\left\langle \frac{1}{A_t}\right\rangle-\frac{1}{\langle A_t\rangle}\right).
  \end{align*}
\end{appendix}

\bibliographystyle{custom.bst} 
\bibliography{./bibliography.bib,addon-long.bib}
\end{document}